\documentclass[pra,superscriptaddress,twocolumn]{revtex4}
\usepackage{enumerate}
\usepackage{amsfonts,amssymb,amsmath}
\usepackage[]{graphics,graphicx,epsfig}
\usepackage{amsthm}
\usepackage{float}
\usepackage{graphicx}
\usepackage{dcolumn}
\usepackage{natbib}
\usepackage{color}
\usepackage{multirow}
\usepackage{ulem}
\usepackage{float}
\usepackage{svg}
\usepackage{lineno,hyperref}
\usepackage{color}
\usepackage{amsfonts,amssymb,amsmath}
\modulolinenumbers[5]

\begin{document}

\title{Electrically coupled optomechanical cavities as a tool for quantum nondemolition measurement}

\author{Jan W{\'o}jcik}
\affiliation{Institute of Spintronics and Quantum Information, Faculty of Physics, Adam Mickiewicz University, 61-614 Pozna\'n, Poland}
\author{Grzegorz Chimczak}
\email{grzegorz.chimczak@amu.edu.pl}
\affiliation{Institute of Spintronics and Quantum Information, Faculty of Physics, Adam Mickiewicz University, 61-614 Pozna\'n, Poland}

\begin{abstract}
    We present a new model of two electrically coupled optomechanical cavities. This model is based on the recently presented [Physical Review A \textbf{103} (2021) 043509]. We found that coupling two optomechanical cavities via Coulomb force leads to cross-Kerr interactions between those cavities. We show that such systems may be ideal for a protocol of quantum non-demolition measurement because it is easy to eliminate the self-phase modulation effect. Moreover, nonlinearities in our model are based on easily adjustable parameters, and therefore, given recent experimental studies, we believe that experimental realization of a cross-Kerr interaction via Coulomb force coupling is feasible. 
\end{abstract}

\maketitle

\section{Introduction}
The field of cavity optomechanics has been greatly explored in recent decades. Optomechanics covers the interaction between electromagnetic field and mechanical motion. Review of that field was greatly done by~\cite{Aspelmeyer:2014vr} and as they pointed out optomechanical couplings were found to be useful in various experiments. Recently, these studies have been extended to optomechanical cavities connected electrically to a charged body~\cite{Zhang:2012aa,Xiong:2017tz,Xiong:2017uk,Feng:2021vg}. 
As the authors of Ref.~\cite{Feng:2021vg} shown, such systems can provide nonlinearities described by Hamiltonians with a term proportional to the square of $n$ (photon number operator). Such a term leads to a nonlinear spectrum, thus allowing for observation of the photon blockade effect. A number of experiments and proposals were made exploiting optomechanics and nonlinearity~\cite{Agarwal:2010vt,Groblacher:2009vb,Hensinger:2005vw,LaHaye:2004vr,Sankey:2010ur,Thompson:2008wk,Zhang:2012aa,Abo_2022,Lai22_tri}. There are many other phenomena that can be observed when non-linearity is present in optical systems~\cite{Kowalewska19, Kalaga19, Kostrzewa22}. An example of that is a quantum nondemolition measurement (QND), which has been studied for a few decades now and a variety of protocols have been presented on that topic. This phenomenon makes it possible to count photons without absorption, and thus, QND is still a rapidly explored field \cite{Gu:2017ut,Jacobs:1994uc,Niemietz:2021ur,Pinard:1995tv,Raimond:2001ts,Reiserer:2015uz}. However, as was pointed out by Balybin \textit{et al.}~\cite{Balybin:2022uc}, schemes which has been proposed to realize QND up to now, are very complicated and experimentally challenging. Over the last decade mostly atom-based QND schemes were developed. Most of the schemes discovered up to now are not ideal for QND in the sense that, apart from the product of photon number operators of different cavities $n_1 n_2$ in a Hamiltonian, which is the soul of QND, there is also a term proportional to the square of photon number operator. This leads to the self-phase modulation effect \cite{Imoto:1985ua,Balybin:2022uc}, which is an obstacle and has to be taken into account when performing QND measurements. 

Here, we propose to engineer a nonlinear cross-Kerr interaction between two optomechanical cavities by giving electrical charge to their movable mirrors. We also prove that it is possible to eliminate the unwanted self-phase modulation effect by adding two charged bodies on both sides of this device. Therefore, we argue that electrically coupled optomechanical cavities can be perfectly suitable for QND.

\section{Model}
\begin{figure*}[t]
  \centering
  \includegraphics[width=17cm]{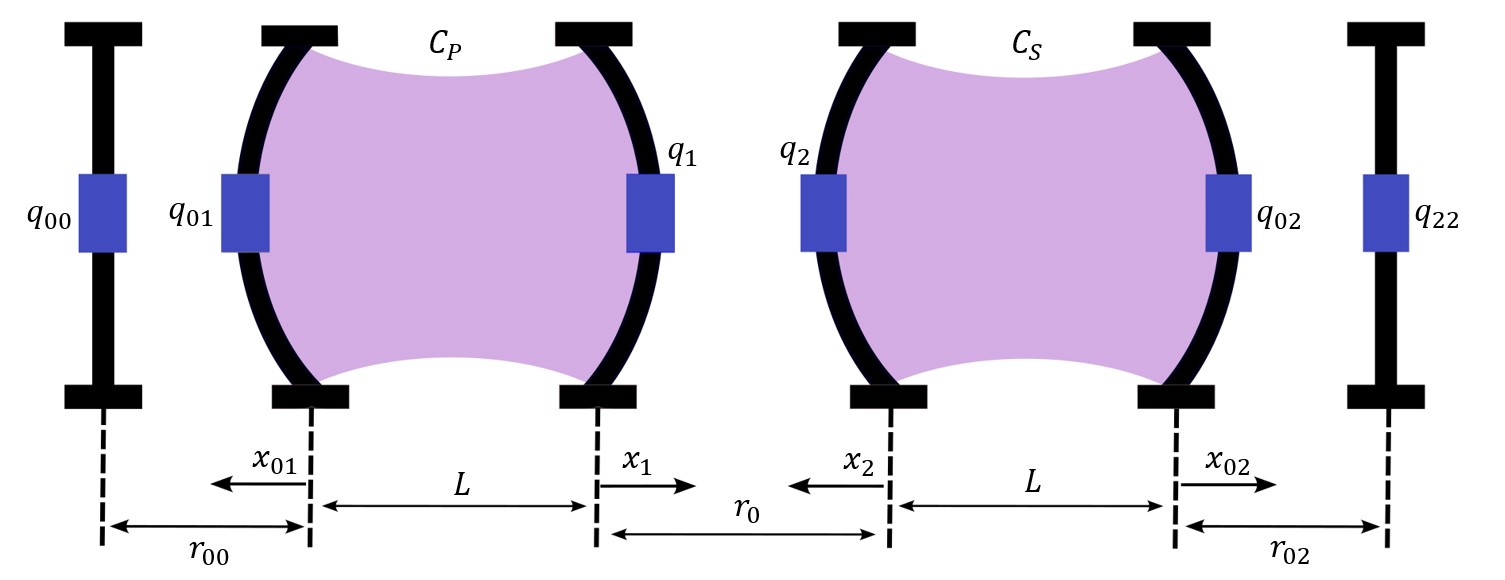}
  \caption{Sketch of the optomechanical setup to perform QND without the self-phase modulation effect. The contribution to this unwanted effect from the electric charges $q_1$ and $q_2$ is compensated for by the contribution of the outside electric charges $q_{00}$ and $q_{22}$.}
  \label{fig:model}
\end{figure*}

Our scheme consists of two optomechanical cavities (probe cavity $C_P$ and signal cavity $C_S$) coupled by the Coulomb force to the charged bodies similarly to the scheme proposed in Ref.~\cite{Feng:2021vg}, but with extra coupling between the two cavities. This setup is shown in Fig.~\ref{fig:model}.
Coupling between cavities is also Coulombian and crucial for obtaining cross-Kerr interactions. Hamiltonian of our system is given by
\begin{eqnarray}
  H &=& H_0 + H_{\rm{om}}+H_{co}\, ,
\end{eqnarray}
where
\begin{eqnarray}
    H_0 &=& \hbar\omega_c a_1^\dagger a_1   +   \hbar\omega_c a_2^\dagger a_2 \nonumber\\
    &&+\frac{m}{2}\omega_{m}^2 x_{01}^2 + \frac{p_{01}^2}{2m}+  \frac{m}{2}\omega_{m}^2 x_{1}^2 + \frac{p_{1}^2}{2m}\nonumber\\
    &&+\frac{m}{2}\omega_{m}^2 x_{2}^2 + \frac{p_{2}^2}{2m}+  \frac{m}{2}\omega_{m}^2 x_{02}^2 + \frac{p_{02}^2}{2m}
\end{eqnarray}
describes the energy of cavities and mechanical oscillators without any interactions between them,
\begin{eqnarray}
    H_{om} &=& -\hbar g_0 (a_1^\dagger a_1 x_{01} + a_1^\dagger a_1 x_1+a_2^\dagger a_2 x_2+a_2^\dagger a_2 x_{02})\nonumber
\end{eqnarray}
describes the coupling between mechanical modes and cavities and
\begin{eqnarray}
    H_{co} &=& H_{\rm{co1}}+H_{\rm{co0}}
\end{eqnarray}
describes Coulombian interactions with
\begin{eqnarray}
    H_{co1} &=& \frac{k q_1 q_2}{r_{0} + x_2-x_1}\, ,\nonumber\\
    H_{co0} &=& \frac{k q_{01} q_{00}}{r_{00} +x_{01}}+\frac{k q_{02} q_{22}}{r_{02} -x_{02}}\, ,
\end{eqnarray}
where $a_{1,2}$ ($a_{1,2}^\dagger$) are annihilation (creation) operators for the first and the second optical mode, respectively, and $x$, $p$ are position, momentum operators for mechanical oscillator modes with frequency $\omega_m$ and mass $m$, $g_0 = \omega_c/L$ is the coupling strength between cavity of length $L$ and mechanical oscillator.\\
For simplicity reasons we introduce new notation and we assume that parameters of the system are symmetric
\begin{itemize}
    \item $k(q_{01}\cdot q_{00}) = k(q_{22} \cdot q_{02}) = \rho_0$
    \item $k(q_1 \cdot q_2) = \rho$
    \item $r_{02} = r_{00} = R_0$
\end{itemize}
To justify omitting interactions between further charges, we assume that closest charges are of different sign $\rho_0<0$, $\rho <0$ and that both $r_0$ and $R_0$ are much smaller then $L$. 

\section{Effective Hamiltonian} 
To deal with Hamiltonian $H$ first we expand its Coulombian part $H_{co}$ to second order of $x/r_0$
\begin{eqnarray}
    H_{co1} &=& \frac{\rho}{r_{0} + x_2-x_1} \nonumber\\
    &\approx& V_{0} +\frac{\rho}{r_{0}^2}(x_1-x_2)+\frac{\rho}{r_{0}^3}(x_1-x_2)^2
\end{eqnarray}
and then we shift the equilibrium point by introducing new position operators $\Tilde{x}_1 =x_1  - d_1$, and $\Tilde{x}_2 =x_2 - d_2$, where $d_1 = -d_2 = -\alpha r_0 / 2(m\omega_m ^2 r_0 +2\alpha)$, $\alpha = k q_1 q_2 / r_0^2$.
Note that one can find $d_1$ and $d_2$ by calculating the minimum of potential
\begin{eqnarray}
    V&=&\frac{\rho}{r_{0} + x_2-x_1} +\frac{m}{2}\omega_{m}^2 (x_{1}^2+x_{2}^2)\, .
\end{eqnarray}
Finally, we obtain
\begin{eqnarray}
    H_{co1} &=& -Q(\Tilde{x}_1-\Tilde{x}_2)^2\, ,
\end{eqnarray}
where $Q = -\rho /r_0^3$ and same we do for $H_{co0}$, and we get 
\begin{eqnarray}
    H_{co0} &=& -Q_0(\Tilde{x}_{01}^2+\Tilde{x}_{02}^2)
\end{eqnarray}
with $Q_0=-\rho_0 /R_{0}^3$ and $\Tilde{x}_{01} =x_{01}  - d_{01}$, and $\Tilde{x}_{02} =x_{02} - d_{02}$, where
\begin{eqnarray}
    d_{01} &=& k q_{00} q_{01} /2(m\omega_m ^2 r_{00}^2 +k q_{00} q_{01} / r_{00})\, ,\nonumber\\
    d_{02} &=& -k q_{22} q_{02} /2(m\omega_m ^2 r_{00}^2 +k q_{22} q_{02} / r_{00})\, .
\end{eqnarray}
Now, using position and momentum operators in the form
\begin{eqnarray}
    x&=&\sqrt{\frac{\hbar}{2m\omega_m}}(b^\dag+b)\, ,\nonumber\\
    p&=&i\sqrt{\frac{\hbar m\omega_m}{2}}(b^\dag-b)\, ,
\end{eqnarray}
with $b_{i}$ ($b_{i}^\dagger$) being annihilation (creation) operators for mechanical modes, we can rewrite $H$ as
\begin{equation}
\label{eq:H_in_terms_of_b}
    H = \hbar\Delta_1 a_1^\dagger a_1   +   \hbar\Delta_2 a_2^\dagger a_2 + H_I + H_Q,
\end{equation}
with $H_I$ describing interaction between cavities via charges $q_1$ and $q_2$, and $H_Q$ describing interactions between cavities and charged bodies $q_{00}$ and $q_{22}$:
\begin{eqnarray}
    H_I &=& \hbar(\omega_m - 2G)(b_1^\dagger b_1 + b_2^\dagger b_2)\nonumber\\
    &&-\hbar g a_1^\dagger a_1 (b_1^\dagger + b_1) - g a_2^\dagger a_2 (b_2^\dagger + b_2)\nonumber\\
    &&-\hbar G\left(\right. b_1^{\dagger2}+ b_1^2 + b_2^{\dagger 2}+ b_2^2\nonumber\\
    &&- 2(b_1^\dag+b_1)(b_2^\dag+b_2)\left.\right)\, ,\nonumber\\
    H_{Q} &=& \hbar(\omega_m - 2G_0)(b_{01}^\dagger b_{01} + b_{02}^\dagger b_{02})\nonumber\\
    &&- \hbar g a_{01}^\dagger a_{01} (b_{01}^\dagger + b_{01}) \nonumber\\
    &&- \hbar g a_{02}^\dagger a_{02} (b_{02}^\dagger + b_{02})-\hbar G_0\left(\right. b_{01}^{\dagger2}+ b_{01}^2 \nonumber\\
    &&+ b_{02}^{\dagger 2}+ b_{02}^2\left.\right)\, ,
\end{eqnarray}
where $\Delta_1 = \omega_c -g_0(d_1+d_{01})$, $\Delta_2 = \omega_c -g_0(d_2+d_{02})$, $G = Q/(2m\omega_m)$, $G_0 = Q_0/(2m\omega_m)$ and $g = \omega_c\sqrt{\hbar/2m\omega_m}/L$. 
\begin{figure*}
\centering
\includegraphics[width=17cm]{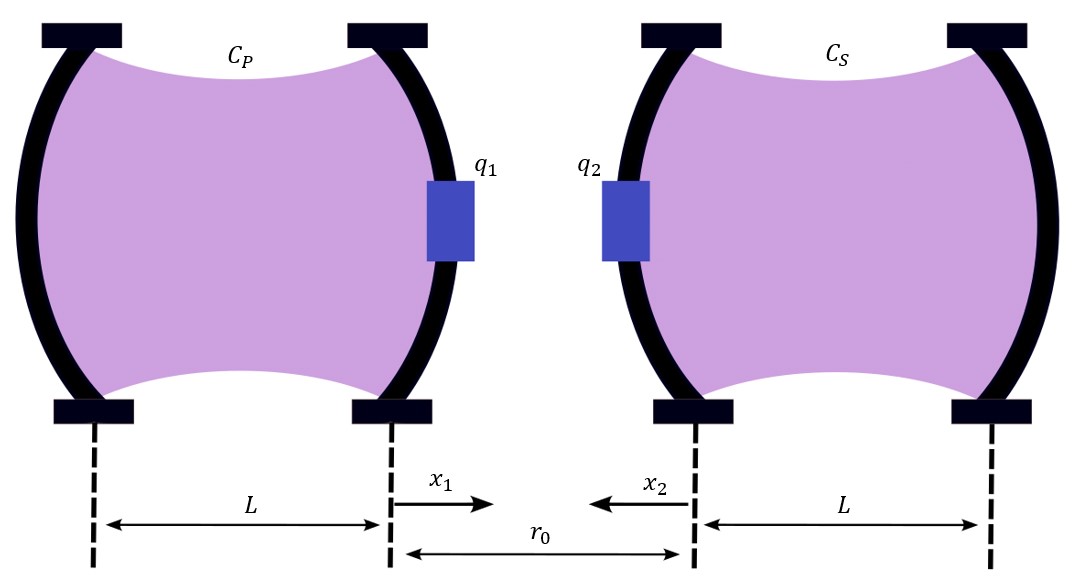}
\caption{The basic version of the setup to perform QND measurement via a cross-Kerr interaction. In this system the unwanted self-phase modulation effect is present.}
\label{fig:simple}
\end{figure*} 
It is possible to simplify the Hamiltonian~(\ref{eq:H_in_terms_of_b}) by applying adiabatic eliminations of all mechanical modes. To this end, we first transform $H_{Q}$ by introducing squeezed mechanical oscillator modes (see \cite{Feng:2021vg} and Appendix A). Then we can eliminate adiabatically these squeezed mechanical oscillator modes provided that $\omega_s\gg g_s$(see Appendix A). After these transformations we get
\begin{eqnarray}
    H_{Q\textrm{eff}} = \hbar g^2\frac{\omega_m}{\omega_m(\omega_m-4G_0)}(n_1^2+n_2^2)\, ,
\end{eqnarray}
where $n_{i} = a_{i}^\dagger a_{i}$ is a photon number operator for the $i$-th mode.\\
Now we perform similar transformations on $H_I$. To eliminate adiabatically both mechanical modes properly, we first diagonalize the bare mechanical part of the Hamiltonian:
\begin{eqnarray}
    H_{m} &=&\hbar (\omega_m-2 G) (b_1^\dagger b_1+b_2^\dagger b_2)\nonumber\\
    &&-\hbar G\left( b_1^{\dagger2} + b_1^2 + b_2^{\dagger 2} + b_2^2 - 2(b_1^\dag+b_1)(b_2^\dag+b_2)\right)\, .\nonumber
\end{eqnarray}
To this end, we assume $\omega_{m}>8\,G$ and use a Hopfield-Bogoliubov transformation:
\begin{eqnarray}
    B_1 &=& \frac{\sqrt{\nu+1}}{2} b_1 -\frac{\sqrt{\nu-1}}{2}\, b_1^{\dagger}  
     -\frac{\sqrt{\nu+1}}{2} b_2 + \frac{\sqrt{\nu-1}}{2} b_2^{\dagger}\, ,\nonumber\\
    B_2 &=& \frac{1}{\sqrt{2}} b_1 + \frac{1}{\sqrt{2}} b_2\, ,
\end{eqnarray}
where $\lambda_1=\sqrt{\omega_{m} (\omega_{m} -8\,G)}$, $\lambda_2=\omega_{m}$ and $\nu=(\lambda_2 - 4\,G)/\lambda_1$. It can be checked that these operators satisfy the canonical commutation relations $[B_i,B_j^{\dagger}]=\delta_{i,j}$. In terms of these operators, we can express $H_{m}$ in diagonal form:
\begin{eqnarray}
    H_{m} &=& \hbar \lambda_1 B_1^{\dagger} B_1 
    + \hbar\lambda_2 B_2^{\dagger} B_2 - \hbar\chi\, ,
\end{eqnarray}
where $\chi=(\omega_{m}-4\,G-\sqrt{\omega_{m} (\omega_{m}-8\, G)})/2$. It is also necessary to express the operators $b_1$ and $b_2$ in terms of $B_1$ and $B_2$:
\begin{eqnarray}
b_1&=& \frac{1}{\sqrt{2}} B_2+\frac{\sqrt{\nu +1}}{2} B_1+\frac{\sqrt{\nu -1}}{2} B_1^{\dagger}\, ,\nonumber\\
b_2&=& \frac{1}{\sqrt{2}} B_2-\frac{\sqrt{\nu +1}}{2} B_1-\frac{\sqrt{\nu -1}}{2} B_1^{\dagger}\, ,
\end{eqnarray}
Now we can re-express $H_I$ to the form
\begin{eqnarray}
\label{eq:HB1B2}
    H_I &=&\hbar \lambda_1 B_1^{\dagger} B_1 + \hbar\lambda_2 B_2^{\dagger} B_2 - \hbar\chi\nonumber\\
    &&-\hbar\frac{g}{2}(\sqrt{\nu-1}+\sqrt{\nu+1}) (a^{\dagger}_{1} a_{1}-a^{\dagger}_{2} a_{2}) (B_1^\dagger+B_1)\nonumber\\
    &&-\hbar\frac{g}{\sqrt{2}} (a^{\dagger}_{1} a_{1}+a^{\dagger}_{2} a_{2}) (B_2^\dagger+B_2)\, .
\end{eqnarray}
After adiabatic elimination shown in Appendix B one obtains
\begin{eqnarray}
\label{HJ}
H_{I\textrm{eff}} &=& \hbar g^2\,\frac{8G}{\omega_m(\omega_m-8G)}\,(n_1 \, n_2)\nonumber\\
    &&-\hbar g^2\,\frac{\omega_m-4G}{\omega_m(\omega_m-8G)}\,(n_1^2 + n_2^2)\, .
\end{eqnarray}
By combining the above results we get
\begin{eqnarray}
   \label{eq:Hcomb}
    H_{\textrm{eff}} &=& \hbar\Delta_{1} a^{\dagger}_{1} a_{1} +\hbar\Delta_2 a_2^\dagger a_2
    +\hbar g^2\,\frac{8G}{\omega_m(\omega_m-8G)}\,(n_1 \, n_2)\nonumber\\
    &&-\hbar g^2\,\frac{\omega_m-4G}{\omega_m(\omega_m-8G)}\,(n_1^2 + n_2^2)\nonumber\\
    &&+\hbar g^2\frac{\omega_m}{\omega_m(\omega_m-4G_0)}(n_1^2 + n_2^2)
\end{eqnarray}
One can see that the last two terms in Eq.~(\ref{eq:Hcomb}) describe the self-phase modulation effect. The first of these terms depends on charges $q_{01}$, $q_{00}$, $q_{22}$ and $q_{02}$, while the second depends on charges $q_{1}$ and $q_{2}$. Since these two terms have different signs, it is possible to eliminate this unwanted in QND effect just by setting proper values of the charges $q_{01}$ and $q_{00}$. The proper values of $q_{01}$, $q_{00}$ (and thus also $q_{22}$ and $q_{02}$) can be determined using the condition
\begin{eqnarray}
  G_0 &=& \frac{\omega_m G}{\omega_m - 4G}\, .
\end{eqnarray}
After setting proper values of these charges the effective Hamiltonian, which is ideal for QND measurements \cite{Imoto:1985ua}, takes the form
\begin{eqnarray}
\label{eq:Heffbest}
    H_{\textrm{eff}} &=& \hbar\Delta_{1} n_1 +\hbar\Delta_2 n_2+\hbar\gamma(n_1 \, n_2)\, ,
\end{eqnarray}
where
\begin{eqnarray}
    \gamma &=& g^2\,\frac{8G}{\omega_m(\omega_m-8G)}\, .
\end{eqnarray}
To sum up let us collect all the conditions that must be satisfied for the effective Hamiltonian~(\ref{eq:Heffbest}) to correctly describe the system shown in Fig.~\ref{fig:model}. These conditions are given by: $\omega_m>8G$, $\omega_s\gg g_s$, $r_0\ll L$, $r_{00}\ll L$ and $G_0 =\omega_m G/(\omega_m - 4G)$. Given similar experimental setups, we believe that these conditions can be met~\cite{Agarwal:2010vt,Groblacher:2009vb,Hensinger:2005vw,LaHaye:2004vr,Sankey:2010ur,Thompson:2008wk,Zhang:2012aa}. 

\section{Simplified model}
The setup proposed in the previous section allows for QND measurements without the self-phase modulation effect. However, the experimental realization of this system might be challenging. Therefore, we also propose a simpler version of the system which also makes it possible to perform QND measurements via a cross-Kerr interaction. This simplified system is shown in Fig.~\ref{fig:simple}. The Hamiltonian of this system is given by
\begin{eqnarray}
    H &=& \hbar\Delta_{1}' a^{\dagger}_{1} a_{1} +\hbar\Delta_2' a_2^\dagger a_2
    + \hbar(\omega_m - 2G)(b_1^\dagger b_1 + b_2^\dagger b_2)\nonumber\\
    &&- \hbar g a_1^\dagger a_1 (b_1^\dagger + b_1) -\hbar g a_2^\dagger a_2 (b_2^\dagger + b_2)\nonumber\\
    &&-\hbar G\left(\right. b_1^{\dagger2}+ b_1^2 + b_2^{\dagger 2}+ b_2^2 - 2(b_1^\dag+b_1)(b_2^\dag+b_2)\left.\right)\, ,\nonumber
\end{eqnarray}
where $\Delta_1' = \omega_c+g_0 d_1$, $\Delta_1' = \omega_c+g_0 d_2$. Following our previous steps, we get effective Hamiltonian in the form
\begin{eqnarray}
\label{eq:Heff_s}
    H_{\textrm{eff}} &=& \hbar\Delta_{1}' n_{1} +\hbar\Delta_2' n_2
    +\hbar g^2\,\frac{8G}{\omega_m(\omega_m-8G)}\,(n_1 \, n_2)\nonumber\\
    &&-\hbar g^2\,\frac{\omega_m-4G}{\omega_m(\omega_m-8G)}\,(n_1^2 + n_2^2)\, .
\end{eqnarray}
The above effective Hamiltonian has a cross-Kerr interaction term, and thus, it also can be used for QND but with so called self-phase modulation effect described in Ref.~\cite{Imoto:1985ua}.

\begin{figure}[t]
\centering
\includegraphics[width=9cm]{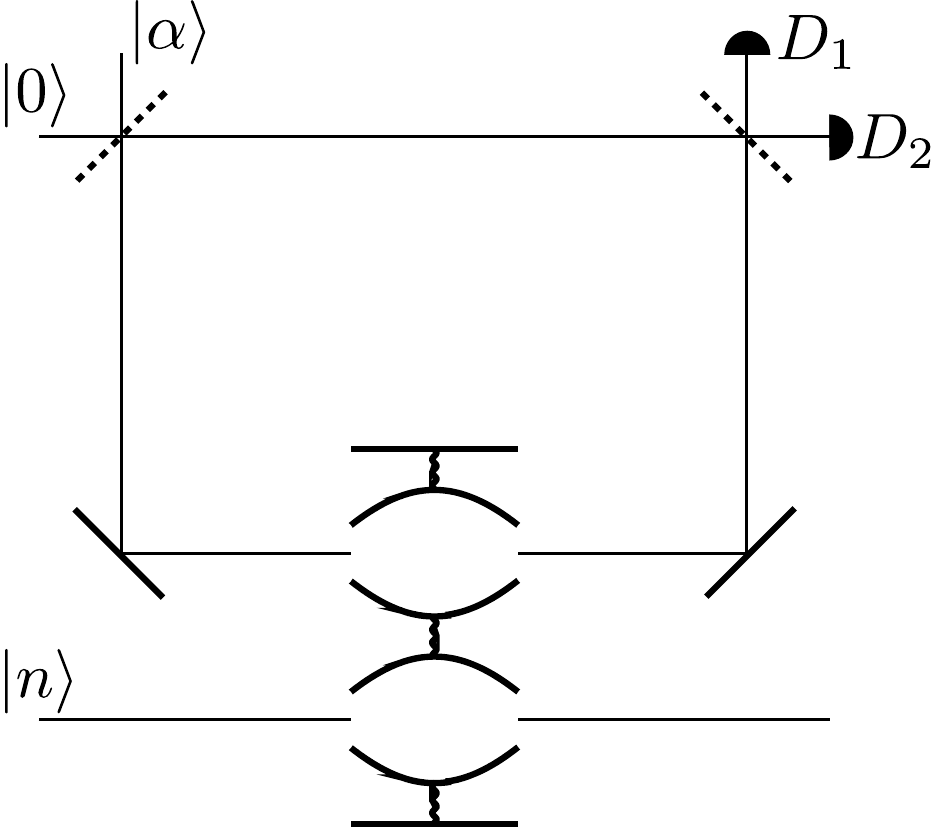}
\caption{Schematic representation of the protocol for a QND measurement using the setup presented in Fig.~\ref{fig:model}. This protocol is simple only if the self-phase modulation effect is absent.}
\label{sh}
\end{figure} 
\section{QND protocol}
Let us now illustrate how to use the setup presented in Fig.~\ref{fig:model} to perform a QND measurement. The protocol for this measurement is depicted schematically in Fig.~\ref{sh}. This protocol exploits a typical phase-shift measurement using a Mach–Zehnder interferometer with coherent states of light. The effect of MZI on coherent states is given, for example, in Ref.~\cite{gerry_knight_2004}. We want to measure $n$, i.e., the number of photons in a signal mode, without destroying it. Initially, the signal mode is prepared in the $|n\rangle$ Fock state, and the two other modes are prepared in the vacuum state and the coherent $|\alpha\rangle$ state, respectively. Therefore, the initial state of the system is given by
\begin{equation}
    |\Psi_0\rangle = |n\rangle_1 |0\rangle_2 |\alpha\rangle_3.
\end{equation}
Firstly, the coherent light falls on the first beam splitter resulting in
\begin{equation}
    |n\rangle_1 |0\rangle_2 |\alpha\rangle_3 \longrightarrow |n\rangle_1 |i\alpha/\sqrt{2}\rangle_2 |\alpha/\sqrt{2}\rangle_3.
\end{equation}
Then, for a time $T$ the state in the lower path (the mode 2) interacts with the signal state (the mode 1) due to the interaction described by the Hamiltonian~(\ref{eq:Heffbest})
\begin{gather}
    |n\rangle_1 |i\alpha/\sqrt{2}\rangle_2 |\alpha/\sqrt{2}\rangle_3 \longrightarrow \nonumber\\ \longrightarrow e^{-i n \Delta_1 T}|n\rangle_1 |ie^{i\theta}\alpha/\sqrt{2}\rangle_2 |\alpha/\sqrt{2}\rangle_3,
\end{gather}
where 
\begin{equation}
\label{eq:theta}
    \theta = -T(\Delta_2+\gamma n).
\end{equation}
Next, the beams interfere at the second beam splitter, resulting in
\begin{gather}\nonumber
    e^{-i n T\Delta_1}|n\rangle_1 |ie^{i\theta}\alpha/\sqrt{2}\rangle_2 |\alpha/\sqrt{2}\rangle_3 \longrightarrow\\
    \longrightarrow e^{-i n T\Delta_1}|n\rangle_1 |i(e^{i\theta}+1)\alpha/2\rangle_2 |(e^{i\theta}-1)\alpha/2\rangle_3 \, .
\end{gather}
Let us denote by $d_1$ ($d_1^\dagger$) and $d_2$ ($d_2^\dagger$) anihilation (creation) operators describing modes collected by detectors $D_1$ and $D_2$, respectively. In the last step, we measure the expectation value of the number difference operator defined by $D = d_1^\dagger d_1 - d_2^\dagger d_2$. It is easy to check that this expectation value is given by
\begin{equation}
\label{eq:expD}
    \langle D \rangle = |\alpha|^2 \cos\theta\, .
\end{equation}
Therefore, if we know the measurement outcome $\langle D \rangle$ then from Eqs.~(\ref{eq:theta}) and (\ref{eq:expD}), we can determine the number of photons in the signal mode
\begin{equation}
    n=\frac{\arccos\left(\frac{\langle D \rangle}{|\alpha|^2}\right)/T - \Delta_2}{\gamma}.
\end{equation}
Thus, we indeed measured $n$ without destroying the signal mode.

It is worth to note that this protocol is simple, at least in theory, thanks to the absence of terms proportional to square of photon number operators in the Hamiltonian~(\ref{eq:Heffbest}). In this case, the time evolution operator $\exp(-i H_{\textrm{eff}} T)$ just transforms one coherent state into another coherent state. However, in cases where terms proportional to operators $n_1^2$ and $n_2^2$ are present in a Hamiltonian, like in the Hamiltonian~(\ref{eq:Heff_s}), the situation is much more complex. Then, the time evolution operator includes the nonlinear Kerr-type operator $\exp(-i \chi  \hat{n}^2)$, which significantly changes a coherent state. Even in special cases, the action of the Kerr-type operator on a coherent state results in the transformation of it into a superposition of many coherent states~\cite{jeong2005}.

\section{Conclusions} 
We have proposed a new setup, in which electrical coupling of two
optomechanical cavities leads to cross-Kerr interactions between them, and therefore, this setup can serve as a quantum non-demolition (QND) measurement device. Moreover, we have also proposed a second version of this setup, in which both cavities interact not only with each other but also with charged bodies. We have shown that the contribution from this additional interaction to self-phase modulation terms can compensate for the contribution to this terms from interactions between both cavities, without changing cross-Kerr interactions terms. Therefore, the effective Hamiltonian of the modified setup includes a cross-Kerr interaction term, which plays a key role in QND measurements, but does not include self-phase modulation terms, which is an obstacle in QND. Finally, we have presented a simple protocol using the modified setup to show how helpful is the eliminating self-phase modulation effect in QND measurements.

\section*{Acknowledgements}
This work was supported by the Polish National Science Centre (NCN) under the Maestro Grant No. DEC-2019/34/A/ST2/00081.

\section*{Appendix A} 
We transform the Hamiltonian $H_Q$ describing interaction between cavities and charged bodies by introducing operators $b_s$ and $b_s^\dag$~\cite{Feng:2021vg}
\begin{equation}
    b=\cosh(r)b_s+\sinh(r)b_s^\dag,
\end{equation}
which satisfy the canonical commutation relation $[b_s, b_s^\dag]=1$. We also set such a value of the squeezing parameter $r$ to fulfill the following condition
\begin{equation}
    r=\frac{1}{4}\log\left[\frac{\omega_m}{\omega_m-4G_0}\right] \, .
\end{equation}
Then, the Hamiltonian $H_Q$ takes the form
\begin{eqnarray}\nonumber
    H_{Q} &=& \hbar\omega_s (b_{s01}^\dagger b_{s01} + b_{s02}^\dagger b_{s02}) -\hbar g_s a_1^\dagger a_1 (b_{s01}^\dagger +b_{s01})\\
    &&-\hbar g_s a_2^\dagger a_2 (b_{s02}^\dagger +b_{s02} )\, ,
\end{eqnarray}
where
\begin{eqnarray}
    \omega_s &=& (\omega_m - 4 G_0)\, \exp(2r)\, ,\nonumber\\
    g_s &=& g\, \exp(r)\, .
\end{eqnarray}
Next, we eliminate adiabatically the mechanical mode $b_{s01}$ assuming that $\omega_s\gg g_s$ and
\begin{eqnarray}
\dot{b}_{s01}&=&i\,[H_Q,b_{s01}]\, ,\nonumber\\
\dot{b}_{s01}&=&0\, .
\end{eqnarray}
In the same way, we eliminate adiabatically the mechanical mode $b_{s02}$ obtaining
\begin{eqnarray}
    H_{Q\textrm{eff}} &=& \hbar \frac{g_s^2}{\omega_s}(n_1^2+n_1^2)\, .
\end{eqnarray}
\section*{Appendix B} 
Now, we can apply the adiabatic elimination procedure to the Hamiltonian $H_I$ defined in Eq.~(\ref{eq:HB1B2}). To this end, we derive $B_k$ ($k=1,2$) from the set of equations
\begin{eqnarray}
\dot{B_k}&=&i\,[H_I,B_k]\, ,\nonumber\\
\dot{B_k}&=&0\, .
\end{eqnarray}
Thus
\begin{eqnarray}
B_1=B_1^{\dagger}&=&\frac{g}{2\lambda_1}(\sqrt{\nu-1}+\sqrt{\nu+1}) 
(a^{\dagger}_{1} a_{1}-a^{\dagger}_{2} a_{2}),\nonumber\\
B_2=B_2^{\dagger}&=&\frac{g}{\sqrt{2}\lambda_2} (a^{\dagger}_{1} a_{1}+a^{\dagger}_{2} a_{2})\, .
\end{eqnarray}
Substituting the above expressions for $B_1$ and $B_2$ into Eq.~(\ref{eq:HB1B2}) we get
\begin{eqnarray}\label{HJ2}
H_{I\textrm{eff}} &=& -\hbar\frac{g^2}{4\lambda_1}(\sqrt{\nu-1}+\sqrt{\nu+1})^2 (a^{\dagger}_{1} a_{1}-a^{\dagger}_{2}a_{2})^2\nonumber\\
&&-\hbar\frac{g^2}{2\lambda_2}(a^{\dagger}_{1} a_{1}+a^{\dagger}_{2}a_{2})^2.
\end{eqnarray}
Rearranging the above we obtain
\begin{gather}
\label{HJ3}
H_{I\textrm{eff}} = \hbar g^2\,\frac{8G}{\omega_m(\omega_m-8G)}\,(n_1 \, n_2)\nonumber\\
    -\hbar g^2\,\frac{\omega_m-4G}{\omega_m(\omega_m-8G)}\,(n_1^2 + n_2^2).
\end{gather}

%\section*{References}

%\bibliography{mybibfile}

\end{document}